\documentclass[conference, 10pt]{IEEEtran}
\IEEEoverridecommandlockouts 
\usepackage[]{graphicx}
\usepackage{caption}
\usepackage{subfig} 
\usepackage{amsmath}
\usepackage{amssymb}
\usepackage{epsfig}
\usepackage{cite}
\usepackage{color}
\usepackage{balance}
\usepackage{amsmath}
\usepackage{amsfonts} 
\usepackage{graphicx}
\usepackage{pdfpages}
\usepackage[T1]{fontenc} 
\usepackage{array}
\usepackage{url}
\makeatletter
\g@addto@macro{\UrlBreaks}{\UrlOrds}
\makeatother

\usepackage{color}
\usepackage{wrapfig}
\usepackage{lipsum}
\usepackage[hidelinks]{hyperref}
\usepackage{multirow}
\usepackage{tabularx}
\usepackage{tikz}
\usepackage{nth} 
\usepackage{algpseudocode} 
\usepackage{algorithm,algpseudocode}
\usepackage{breqn}
\usepackage{todonotes}
\usepackage{verbatim}

\begin{document}
\title{Analyzing the Impact of Meteorological\\ Parameters on Rainfall Prediction}
\author{\IEEEauthorblockN{
Muhammad~Salman~Pathan\IEEEauthorrefmark{1},
Jiantao~Wu\IEEEauthorrefmark{1},
Yee Hui Lee\IEEEauthorrefmark{2}, 
Jianzhuo Yan\IEEEauthorrefmark{3}, and
Soumyabrata Dev\IEEEauthorrefmark{1}
}
\IEEEauthorblockA{\IEEEauthorrefmark{1} ADAPT SFI Research Centre, School of Computer Science, University College Dublin, Ireland}
\IEEEauthorblockA{\IEEEauthorrefmark{2} School of Electrical and Electronic Engineering, Nanyang Technological University (NTU), Singapore}
\IEEEauthorblockA{\IEEEauthorrefmark{3} Ministry of Education Faculty of Information Technology, Beijing University of Technology, China}
\thanks{This research has received funding from the European Union's Horizon 2020 research and innovation programme under the Marie Skłodowska-Curie grant agreement No. 801522, by Science Foundation Ireland and co-funded by the European Regional Development Fund through the ADAPT Centre for Digital Content Technology grant number 13/RC/2106_P2.}
\thanks{Send correspondence to S.\ Dev, E-mail: soumyabrata.dev@ucd.ie.}
\vspace{-0.6cm}
}

\maketitle

\begin{abstract}
Rainfall is a climatic factor that affects many human activities like agriculture, construction, and forestry. Rainfall is dependent on various meteorological features and its prediction is a very complex task due to the dynamic climatic nature. A detailed study of different climatic features associated with the occurrence of rainfall should be made in order to understand the influence of each parameter in the context of rainfall. In this paper, we propose a methodical approach to analyze the affect of various parameters on rainfall. Our study uses $5$ years of meteorological data from a weather station located in the United States. The correlation and interdependence among the collected meteorological features were obtained. Additionally, we identified the most important meteorological features for rainfall prediction using a machine learning-based feature selection technique.
\end{abstract}

\IEEEpeerreviewmaketitle

\section{Introduction}
Rainfall plays an important role in the formation of fauna and flora of natural life.  
An accurate prediction of a rainfall event can not only support causal usages but also provide early warnings of floods or traffic accidents\cite{pham2020development}. 
Despite the availability of advanced technology and sufficient amount of weather data, the prediction of rainfall event is extremely complex \cite{manandhar2016gps}. The changing climatic conditions and the increasing greenhouse emissions have made it difficult for humans to properly understand the weather~\cite{fathima2019chaotic}. In order to develop efficient rainfall prediction systems, firstly a thorough study of various meteorological parameters should be made in the context of rainfall occurrence. 


\subsection{Related Work}
There is an extensive research carried out in the area of weather forecasting. 
In \cite{yen2019application}, Principal Component Analysis (PCA) is used to analyze the hourly meteorological data of southern Taiwan. The results showed that the most dominant factors controlling the rainfall are the air pressure and humidity. The authors in \cite{manandhar2019data} proposes a  systematic approach  to  analyze different ground based weather parameters from year 2012-2015 of Nanyang Technological University weather station.  Features such as Precipitable Water Vapor (PWV) and solar radiation stand out for rainfall prediction. Oswal \textit{et al.}~\cite{oswal2019predicting} adopted a pairwise correlation matrix to understand interactions between parameters. 
A correlation-based feature selection technique to assemble an effective subset to develop precipitation prediction model is also presented in \cite{manandhar2018systematic}. 

\subsection{Contributions of the paper}
The main contributions of this paper include:
\begin{itemize}
    \item we present a methodical approach to analyze and assess the impact of different meteorological parameters on the occurrence of rainfall;
    \item we also share the source-code of our methodology in the spirit of reproducible research\footnote{The code related to this paper is available here: \url{https://github.com/Sammyy092/Impact-of-meteorological-parameters-on-rainfall}.}.
\end{itemize}

\section{Results \& Discussions}
\subsection{Dataset}
The rainfall data for our research is collected from National Oceanic and Atmospheric Administration (NOAA) Climate Data Online service (CDO\footnote{https://www.ncdc.noaa.gov/cdo-web/}). 
The downloaded monthly data from Jan 2015-Dec 2020 is recorded from the weather stations located in Alpena Regional Airport, Michigan, U.S. 
The meteorological parameters utilized in our work are: precipitation ($PRCP$), average wind speed ($AWND$), direction of fastest 2-minute wind ($WDF2$), direction of fastest 5-second wind ($WDF5$), fastest 2-minute wind speed ($WSF2$), fastest 5-second wind speed ($WSF5$), average temperature ($TAVG$), minimum temperature ($TMIN$), maximum temperature ($TMAX$). These features were selected because they are interdependent and influence the rainfall. Feature $PRCP$ is rainfall measured in inches (in inches) using a rain gauge, and we consider this feature as the rainfall indicator. In this way we will compare other parameters with $PRCP$ in order to find the relation and influence of each parameter for $PRCP$.


\subsection{Correlation Analysis}
\begin{figure}[htb]
\centering
\includegraphics[width=.45\textwidth]{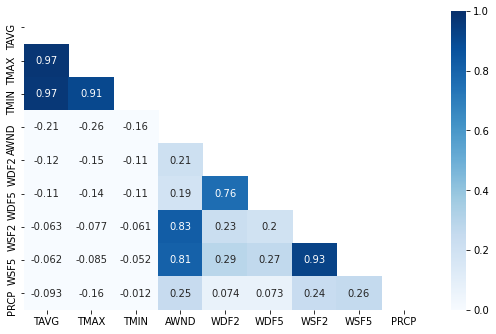}
\caption{We compute the correlation values of the different meteorological parameters and precipitation.}
\label{fig:corr_mat}
\end{figure}

In this section we describe the correlation of different features with respect to $PRCP$. The cross-correlation values among the features were computed by calculating the correlation coefficients of feature matrix $X$, having dimension $m \times n$, denoted as: $X=[v_1, v_2,\ldots, v_n]$, where $v_1, v_2,\ldots, v_n$ are the vectors of $n$ number of meteorological features. Each vector is of length $m$ indicating a weather recording at a particular time. The correlation values of all the features are shown in figure. We observe from Fig.~\ref{fig:corr_mat} that 
there is a significant correlation between $PRCP$ and the features related to data type wind (i.e. $WSF5$, $WSF2$, $WDF5$, $WDF2$, $AWND$). $PRCP$ has a positive correlation with $AWND$, $WSF2$ and $WSF5$ of value (0.25, 0.24 and 0.26).The reason of the positive correlation between the wind and rain is that the winds carry an amount of moisture in it which can highly affect the amount of precipitation in an area. Faster winds and precipitation are strongly correlated in nature where faster winds cause rain, showing major significance on daily rainfall variability.  
Furthermore, using the annual data we found a negative correlation between rainfall and temperature (\textit{i.e.}, as temperature increases, rainfall drops). We can see a value of $-0.167$ when $TMAX$ is related with $PRCP$. The amount of precipitation gets lower with higher temperatures and vice versa. Hence, a strong negative correlation occurs on land, as temperature favor more dry conditions and less evaporative cooling.

\vspace{-0.3cm}

\subsection{Feature Importance}
\begin{figure}[htb]
\centering
\includegraphics[width=.43\textwidth ]{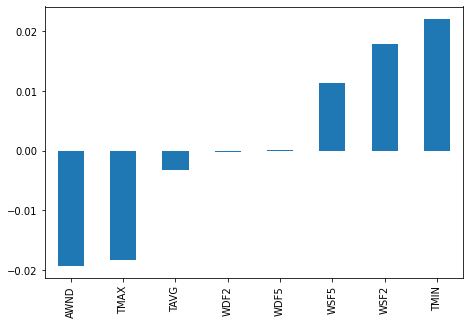}  
\caption{We compute the scores of different features in predicting rainfall.}
\label{fig:us_data}
\end{figure}

This section describes the importance of each input meteorological parameter when predicting the $PRCP$. To accomplish the task, we have used the least absolute shrinkage and selection operator (LASSO)~\cite{li2005lasso}. 
The LASSO is the penalized least squares regression case with $\ell_1$-penalty function which is estimated as:

\[L=\sum_{j}^{}(y_j-\sum_{k}^{}\beta _{k}v_{jk})^2+\lambda \sum_{k}^{} \left \|\beta_{k} \right \|_{1}\]

where $v_{jk}$ denotes the $k$th meteorological feature in the $jth$ datum, $y_{j}$ is the response feature value in this datum and $\beta_{k}$ represents the regression coefficient of the $k$th feature. Because of
$\ell_1$ function $\sum_{k}$ $\left \| \beta _{k} \right \|_{1}$, the LASSO feature selection technique typically produces estimates in which some of the coefficients are set exactly to zero thereby performing 
feature selection. 
Figure~\ref{fig:us_data} shows the feature importance values of each meteorological variable in terms of precipitation prediction. We can observe from the figure that features $TMIN$, $WSF2$, $WSF5$ have the highest values. The weather parameters $TMAX$ and $AWND$ posses the lowest scores meaning that they have a low influence in detecting the rainfall. Furthermore, all the features were selected and given values during the feature selection process, it means that none of the features have a non-zero coefficient after the shrinking process.

\vspace{-0.4cm}

\section{Conclusion \& Future Work}
An accurate prediction of rainfall is not an easy task due to the changing behavior of climate. Therefore, we have presented a methodical analysis of different meteorological parameters associated with the occurrence of rainfall in order to understand the importance of each parameter for rainfall. We obtained 
that 
wind is strongly correlated with rainfall indicating that strong winds can highly influence the occurrence of rainfall. 
We also obtained that wind speed and minimum temperature are 
the most important features in predicting the rainfall. In future we plan to use this reduced subset of features for accurate estimation of the time of occurrence of rainfall.



\bibliographystyle{IEEEtran.bst}

\end{document}